# EUROPEAN LONGITUDE PRIZES. III. THE UNSOLVED MYSTERY OF AN ALLEGED VENETIAN LONGITUDE PRIZE

**Richard de Grijs**
*Department of Physics and Astronomy, Macquarie University,
Balaclava Road, Sydney, NSW 2109, Australia*
Email: richard.de-grijs@mq.edu.au

**Abstract:** Despite frequent references in modern reviews to a seventeenth-century Venetian longitude prize, only a single, circumstantial reference to the alleged prize is known from contemporary sources. Edward Harrison's scathing assessment of the conditions governing the award of an alleged Venetian longitude prize simultaneously disparages the rewards offered by the Dutch States General. However, the latter had long run its course by 1696, the year of the citation, thus rendering Harrison's reference unreliable. Whereas other longitude awards offered by the leading European maritime nations attracted applicants from far and wide, often accompanied by extensive, self-published pamphlets, the alleged Venetian prize does not seem to have been subject to similar hype. The alleged existence of seventeenth-century Venetian award is particularly curious, because the city's fortune was clearly in decline, and longitude determination on the open seas does not appear to have been a priority; the city's mariners already had access to excellent 'portolan' charts. It is therefore recommended that authors refrain from referring to a potentially phantom Venetian longitude prize in the same context as the major sixteenth- to eighteenth-century European awards offered by the dominant sea-faring nations.

## 1 A BANDWAGON EFFECT?

From late-medieval times, major expansion of commercial shipping routes and international trade networks drove significant European efforts to find a practical solution to the perennial 'longitude problem'. Whereas one's latitude could be determined fairly easily based on the altitude of Polaris or that of the Sun's meridian crossing, although adjusted for seasonal variations, determination of one's longitude at sea was not straightforward without access to an accurate marine timepiece (for a recent review, see de Grijs, 2017).

As early as 1567, King Philip II of Castile and Portugal (1527–1598) announced a significant reward for a practical solution to the longitude problem (de Grijs, 2020a). Upon his succession to the throne in 1598, his son—King Philip III of Spain, Portugal and the Two Sicilies (1578–1621)—reaffirmed and expanded the money on offer. Shortly afterwards, on 1 April 1600 the States General of the United Provinces of the Dutch Republic issued a similarly generous award, which was increased step-wise until it had reached 15,000 carolus guilders by 1660 (de Grijs, 2021a). The provincial States of Holland and Westfrisia separately offered a somewhat smaller award as early as 1601.

Both the Spanish and Dutch longitude rewards seem to have run their respective courses by the second half of the seventeenth century. However, by the end of the seventeenth century, the competition for trade opportunities and new colonies had increased significantly among the English, French, Spanish and Dutch, and so developing accurate means of navigation became a matter of national security.

Meanwhile, an increasingly vocal chorus of petitions clamouring for favourable government initiatives reached the authorities in the French Republic and Great Britain, which eventually led to the passing of the well-known British Longitude Act and its associated rewards in 1714 (popularised by Sobel, 1995; see also Andrewes, 1996; Dunn and Higgitt, 2014) and the designation of the French *Prix Rouillé de Meslay* (Meslay Prize; e.g., Boistel, 2015) in the same year. A number of smaller, private initiatives supported by wealthy benefactors and financiers, including Thomas Axe (1635–1691) in Britain (e.g., Turner, 1996; see also de Grijs, 2021b: Paper IV), also jumped onto the bandwagon.

Curiously[1], in most modern scholarly and popular texts on seventeenth- and eighteenth-century efforts to solve the longitude problem, an award allegedly offered by the Venetian Republic is usually included (e.g., in chronological order, Gould, 1923: 12; Usher,

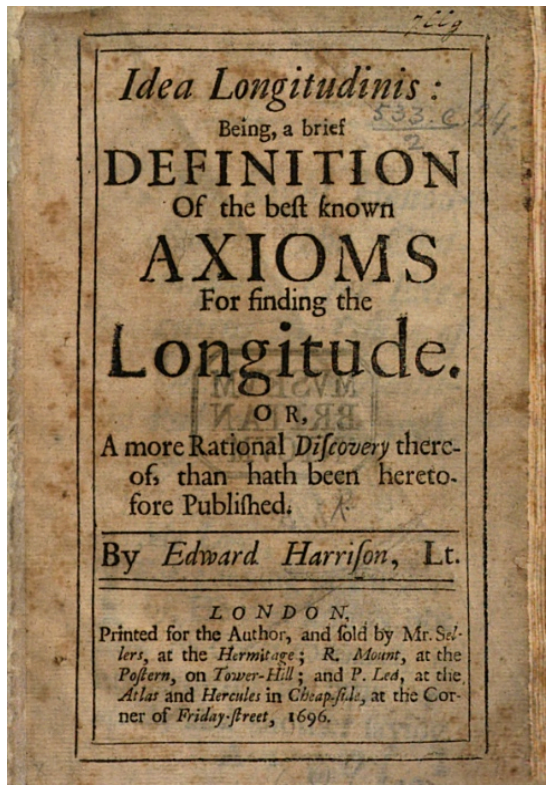

**Figure 1.** Front page of Edward Harrison's treatise, *Idea longitudinis, being a brief definition of the best known axioms for finding the longitude* (London, 1696; out of copyright).

1929: 284; Brown, 1949: 209; Horological Institute of America, 1960: 19; Sobel, 1995; Turner, 1996; Howse, 1998: 26; Galluzzi, 2000; Stephens, 2014; Kochetkova, 2016), although very few authors provide any substance to support the reality of this latter award. At the present time, the only contemporary reference to a Venetian prize comes from Edward Harrison's (fl. 1686–1700) treatise, *Idea longitudinis, being a brief definition of the best known axioms for finding the longitude* (1696; see Figure 1), which was subsequently cited and reinforced by Clark (1764: 19). Harrison (1696: 76–77) talks about a Venetian prize he apparently knows of from hearsay, and he is clearly not impressed; his scathing review of the Axe prize is matched by equally scathing reviews of the Venetian and Dutch rewards (emphasis is mine):

> As I have heard say, **the Venetians have offered Rewards**; the States of Holland (as I was informed,) some Years ago, took this Science so much into their Consideration, that they offered Ten Thousand Pounds, to any Man that was capable to discover it; Thomas Axe, an Englishman, left a Legacy of One Thousand Pound, (never to be paid I think,) to any Person that should discover the Longitude, within the space of Ten Years after his Decease, if his Wife and Child died Childless in that time; besides, it is to be approved of by the four Professors of Geometry and Astronomy, in Oxford and Cambridge, for the time being, and at least twenty Experienc'd Masters of Ships, that shall have made several Experiments thereof in long Voyages. Affidavits are to be made before the Twelve Judges of England, &c. He dyed in the year 1691, and I think took care enough, that the said one Thousand Pound should be Irrecoverable; indeed, **I have little better Opinion of either the Venetian or Dutch Gratuities**; States may be mistaken in their Policies; God Almighty never ordained that this Science should be well known to the World, by fraudulent means, but if their designs be real, (as you may suppose,) where is the Encouragement?

Harrison was a lieutenant in the British Royal Navy; by the time he published his pamphlet proposing a longitude solution, he appears to have had at least ten years of maritime navigation experience on perhaps six Navy ships and two merchant vessels sailing "… to the East". Harrison seems to have been universally disparaging as regards the quality of the navigation equipment available in practice and the methods used for longitude determination (Cook, 1998; Glennie and Thrift, 2009). Nevertheless, he still seemed to have had confidence that a practical approach to longitude determination might be found:

> Some blockheads are apt to say, the Longitude cannot be found: no, it cannot Accidentally ... but by Care, Diligence and Industry; it may be found, without which it cannot be understood. (Harrison, 1696: Dedication)

For all his bluster, he appears to have been fairly well educated. He read Latin, which suggests that he attended grammar school, and he was apparently well versed in the complexities associated with the lunar distance method of longitude determination (for a recent review, see de Grijs, 2020b). Glennie and Thrift (2009) point out that although Harrison did not complete a university degree, he had attended a number of lectures at Gresham College in London, whereas he was also familiar with a range of publications by the Royal Society of London. Yet, despite his education, or perhaps because of it, he preferred to rely on the experience of 'tarpaulins' or 'Jack Tars' (sailors; see Figure 2) rather than on natural philosophers ("scientists").

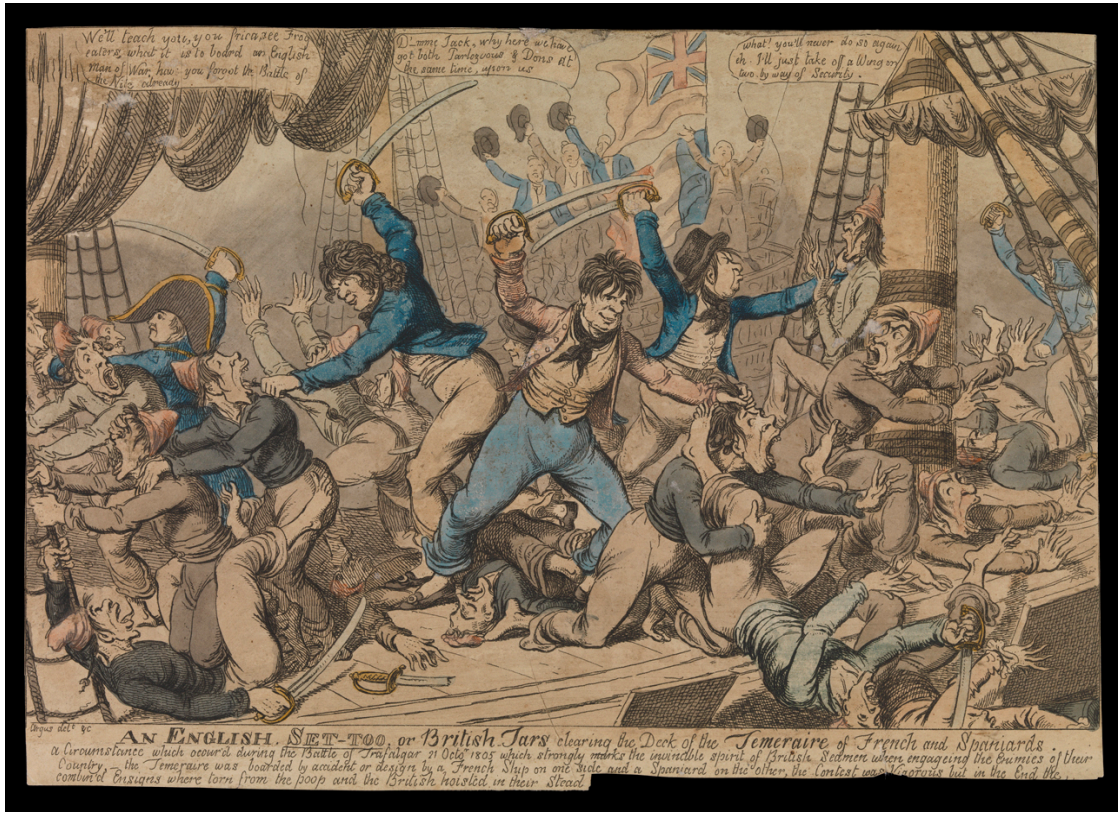

**Figure 2.** British 'Jack Tars', some wearing their typical 'tarpaulin' garments. (© *National Maritime Museum, Greenwich, London. Low-resolution image used with permission; PAF3907*)

However, despite his practical inclination, Harrison's (1696) proposed method of longitude determination fell far short of offering a practical solution:

> Scientifically and technically the book [Harrison's 1696 pamphlet] is poor, second-hand and ill-digested. It is shot through with aspersions on mathematicians, the manner is aggressive, and it shows a deep inferiority complex. (Cook, 1998: 264)

**2 TIMING**

**2.1 Harrison's 1696 claim**

Given that we only have a single, hearsay source for the oft-referenced Venetian longitude prize, can we use other information to constrain the period during which that prize may have been 'live'? Our 'firm' data point is Harrison's 1696 reference, although one should realise that this might be less constraining than it may appear at first sight.

In the same quotation where he disparages the Venetian prize he has ostensibly heard of, he also criticises the Dutch longitude reward—presumably the prize offered by the States General. However, that latter reward had well and truly run its course by 1660,[2] almost four decades prior to the publication of Harrison's pamphlet. As discussed by de Grijs (2017, 2021a), in 1658–1660 Robert Hooke (1635–1703) queried whether his spring-driven timepieces might be eligible for one of the European longitude rewards, but it appears that they had become largely inactive by then (Robertson, 1931: 172). In the Dutch Republic, the nation's East India Company had taken on the role of adjudicator of any methods submitted for consideration for possible compensation, without further reference to the national or provincial awards on offer earlier that century.

Instead, we will now consider submissions for any of the longitude rewards on offer by Venetian hopefuls both before and after the date of Harrison's publication. After all, if the Venetian Republic offered an attractive award for a practical longitude solution, one would expect local[3] scholars and 'projectors' to have submitted their proposals to the Venetian authorities in first instance. As we will see below, we can use this approach to constrain the period during which the alleged Venetian longitude prize may have been active to that between Galileo Galilei's death in 1642 (or shortly thereafter) and the publication of a pamphlet suggesting a new method of longitude determination by Dorotheo Alimari (ca. 1650–ca. 1727) in 1714.

**2.2 Galileo's rise**

Galileo became a force to be reckoned with in the longitude game following his discovery of the regular motions of Jupiter's satellites, published in his short treatise (pamphlet), *Siderius Nuncius* (Galilei, 1610). He realised that their ephemerides might be useful as accurate celestial clock, thus allowing an observer to determine 'absolute' time anywhere on Earth (provided Jupiter was visible, of course).

As early as July 1612, the long-serving administrator to the Florentian de' Medici family, Belisario Vinta (1542–1613), initiated discussions with the Spanish Crown through Count Orso d'Elci (d. 1668)—Tuscan ambassador to Madrid—to negotiate a trade (Gattei, 2017). The Grand Duke of Toscany, Cosimo II de' Medici (1590–1621), was keen to trade directly with the West Indies—located in the Spanish sphere of influence—from his Tuscan home port of Livorno. In return, Cosimo de' Medici offered to disclose Galileo's method of longitude determination and train Spanish sailors in its use. The de' Medici family considered Jupiter's moons their personal property[4] and approached the trade as if they were selling exclusive use rights. Galileo submitted a formal proposal, in which he also offered the Spanish navy written instructions as well as one hundred telescopes that magnified "… forty or fifty times …" (Galluzzi, 2000).

The initial round of negotiations ran until October that year. A second round involving d'Elci, Curzio Picchena (1554–1626) and Giuliano de' Medici[5] lasted from June 1616 until December 1620, but the parties could not agree on the terms of their trade. A major objection that had been raised against Galileo's approach was voiced thus by d'Elci (see also de Grijs, 2020a):

> In order to put your method into practice it is compulsory and necessary first to see the said stars and their aspects. I do not know how this can be done at sea... For, leaving aside the fact that the telescope cannot be used on ships because of their motion, and allowing that it could be used, it could serve neither during the day when the weather is overcast nor at night. (Galluzzi, 2000)

Following technical adjustments by Galileo (for details, see de Grijs, 2020a), a final attempt at reaching an agreement was undertaken by Esaù Del Borgo (d. 1631?), royal chamberlain to Tuscan ambassador Averardo de' Medici (d. 1685?), and Secretary of State Andrea Cioli (1573–1641) in May 1630, but plans were finally shelved by October 1632 (Gattei, 2017).

Meanwhile, Galileo had been aware of the similarly large Dutch longitude reward, but he had wanted to wait until negotiations with the Spanish Crown had concluded before approaching the Dutch authorities. Given the breakdown of those negotiations and reassured by the first-rate scholars and experienced navigators serving on the Dutch adjudication panel (see de Grijs, 2021a), in August 1636 he therefore instructed his Parisian friend Élie Diodati (1576–1661)—a Swiss lawyer from Geneva—to initiate negotiations with the States General of the Dutch Republic (see also de Grijs, 2017; Gattei, 2017). By 1640, these negotiations had also broken down, in part owing to Galileo's poor health.

Nevertheless, Galileo—now blind but still at the top of his mental capabilities—continued to improve his proposal for a practical means of longitude determination at sea. His latest invention, and one that had been included in the negotiations with the Dutch government, involved a stable, isochronous marine timepiece aimed at keeping accurate local time between successive astronomical measurements. It was not meant to solve the problem of taking a 'reference' time to sea for later comparison with the ship's local time, as had already been suggested by the Flemish mathematician Regnier Gemma Frisius (1508–1555) in 1530. Since Galileo had lost his eyesight, he described his ideas for a working pendulum clock (see Figure 3) and an improved pinwheel escapement to ensure regular oscillations to his son Vincenzo in 1641, a year prior to the elder Galilei's death. However, Vincenzo Galilei never managed to construct a working model.

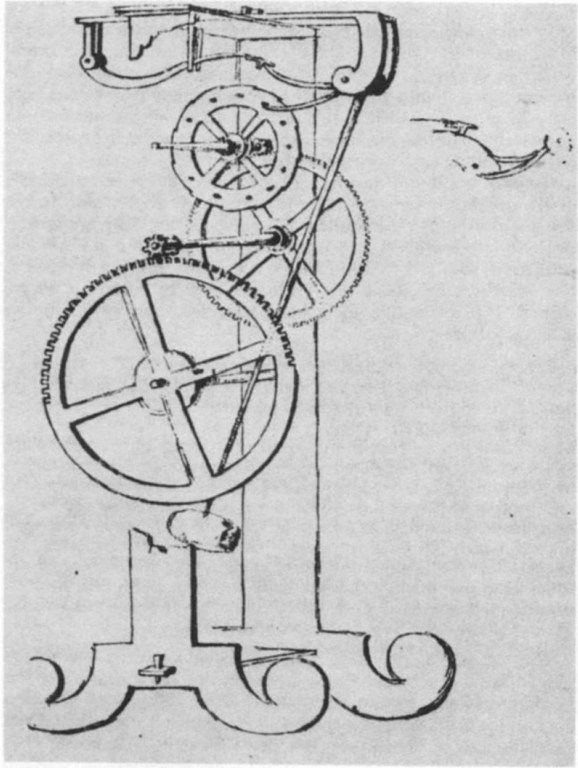

Galileo's decision to negotiate with one party at a time implies that even if he had been aware of a Venetian longitude prize, he may not have wanted to engage with a second, local party simultaneously. Moreover, since he sought the patronage of the Florentine ruling house of de' Medici, it may have been politically self-destructive had he attempted to gain an advantage in Venice as well.

Nevertheless, we know that following Galileo's death in 1642, his last disciple, Vincenzo Viviani (1622–1703), was exceedingly keen to honour his late teacher. Viviani installed a set of inscriptions on the façade of his residence, the Palazzo dei Cartelloni in Florence (see Figure 4). Section D, the longest inscription, celebrates Galileo's achievements, including the use of Jupiter's satellites to determine one's longitude:

**Figure 3.** Galileo's pendulum clock design, drawn by Vincenzo Viviani in 1659. Part of the front supporting plate has been removed by the artist to show the wheelwork. (*Opere di Galileo*, 19, p. 656; *photo: University of Toronto Photographic Services*)

> The satellites of Jupiter, which he discovered first and before all others in Padua, on 7 January 1609, after completing only three observations, he dedicated to the everlasting glory of the Medicean princes. And, due to the observation of their very quick motion, he set forth the question, vainly investigated for a long time, of how to obtain the longitude of places at night, so that under the new auspices of the Medicean House geography and hydrography could be corrected, renewed and refined. And while he worked out over three years of persistent work the periods of the motion of the Medicean Stars and their distances from Jupiter, he prepared canons and tables to predict their very rapidly changing aspects. Disdaining the most generous rewards promised to those who solved such difficult problem, with truly heroic magnanimity he offered his set of theories [of Jupiter's satellites], tables and ephemerides, his optical tubes and his pendulum clock, which he had devised several years earlier in Pisa, and also men skilled in the use of these instruments: first to Philip III, the Catholic King, in 1615, and then to the confederate Provinces of Holland, in 1635. However, by the decree of Omnipotent God the generous offer and noble attempt came to nothing in both cases, so that this essential work for the good of navigation and geography, was begun by means of the Medicean Stars and carried through owing to the liberality of Louis the Great, the most powerful on land and sea, and to the labours of that most excellent astronomer [Giovanni Domenico] Cassini. (Gattei, 2017: 217–218, transl.)

It stands to reason that Viviani would likely have attempted to have Galileo's method of longitude determination heard by the Venetian authorities if such an award were available at the time. As such, and given the absence of any reference to a Venetian longitude prize in Galileo's extensive and well-preserved correspondence, it appears reasonable to suggest that a Venetian longitude prize had not (yet) been established—or was poorly known, even to local scholars—by the mid-1640s.

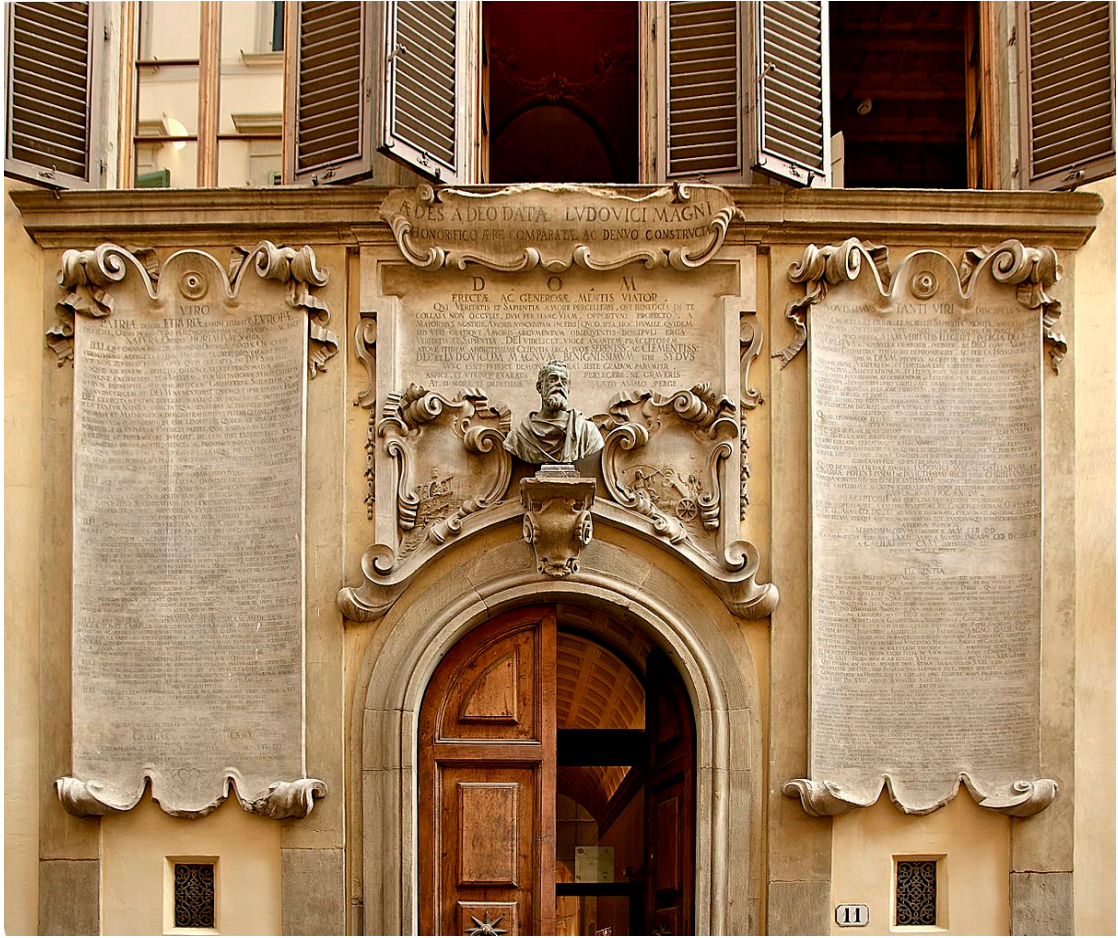

**Figure 4.** Façade, Palazzo dei Cartelloni (Palace of the Placards), 11 via Sant'Antonino, Florence (*Credit: Jebulon via Wikimedia Commons; Creative Commons CC0 1.0 Universal Public Domain Dedication*).

**2.3 Venetian projectors and the British Longitude rewards**

On the other side of the timeline, the first reference to a Venetian proposal purporting to determine longitude at sea dates from 1714. On 31 August of that year, the Venetian painter Sebastiano Ricci (1659–1734) presented one of many pamphlets to the Commissioners of the Longitude in Britain, this one on behalf of the Venetian mathematician, military and naval writer Dorotheo Alimari, titled *The NEW METHOD Propos'd by Sign$^r$ DOROTHEO ALIMARI, Professor of Mathematicks at Venice, to Discover the LONGITUDE … Humbly Presented to Right Honourable … Lords, … Appointed by Act of Parliament COMMISSIONERS, for Examining and Judging of Proposals for Finding the LONGITUDE*.

This new method revolved around an instrument that outwardly resembled both an astrolabe and a sundial (see Figure 5). The poorly printed pamphlet of August 1714 was presented in both Latin and English; an expanded version in Latin—*Longitudinis aut terra aut mari investiganda methodus*—was released early in 1715. The expanded version included images of the instrument, solar tables and a description of a method to predict the tides. Alimari's proposal involved the compilation of an extremely exact almanac for a reference meridian. The user would then compare his observations of the Sun's altitude with the almanac's ephemeris tables. Alimari's idea was dismissed forthwith by the Astronomer Royal as "… speculation and nothing more: the difficulty of reducing it to practice is immense." (Flamsteed, 1714).

Like many of his fellow 'projectors', Alimari had been attracted by the generous British Longitude rewards of 1714, which promised as much as £20,000 for the most accurate tier of longitude solutions. His idea was unworkable, but he was astute in terms of self-aggrandisement. It would seem reasonable, therefore, that if a Venetian longitude prize had been on offer around 1714, Alimari would likely have been in contention—and that status would have come with additional publicity for the method(s) on offer.

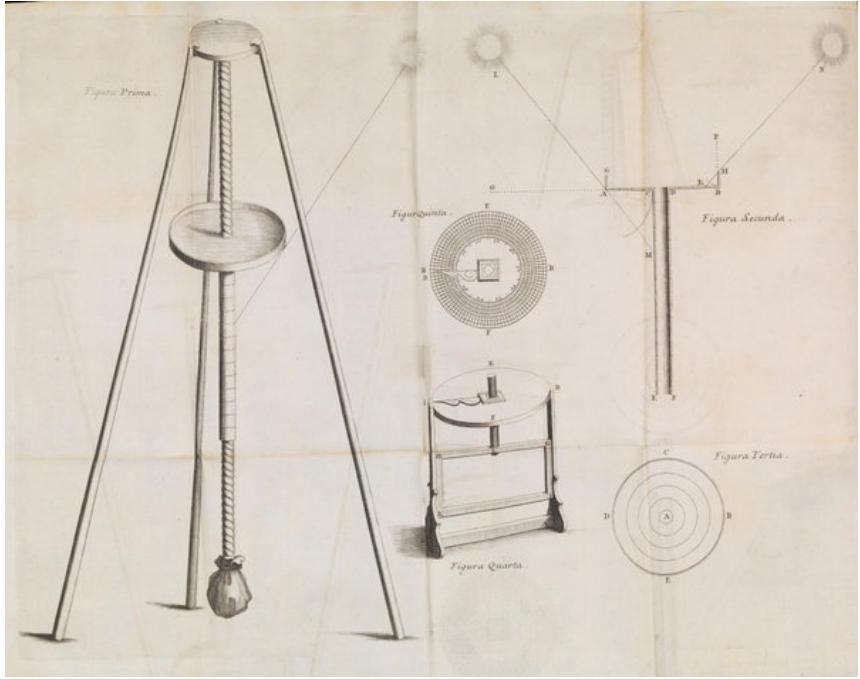

**Figure 5.** Dorotheo Alimari's proposed instrument from his treatise, *The New Method Proposed by Dorotheo Alimari to Discover the Longitude* (© National Maritime Museum, Greenwich, London; Low-resolution image used with permission; L5813-002).

It is indeed odd that not a single formal proposal submitted to the alleged Venetian longitude prize is available in the historical record. The most likely location of any such records would be in the archives of the *Riformatori allo Studio di Padova* (P. Benussi, personal communication; April 2021), which are contained within the State Archives of Venice. The *Riformatori*, the Reformers of the University of Padua, the *Studio*, were three magistrates representing the Republic of Venice at the University in all matters related to education, culture and science. However, their remit was much broader than pertaining to merely matters related to the University.

It is expected that the University of Padua would have been involved in deliberations about any Venetian longitude prize, given the scientific relevance of the subject and the expertise required to evaluate the proposals. However, despite extensive searches and correspondence with scholars in the field, I have been unable to unearth anything of direct relevance to a Venetian longitude prize in these archives.

## 3 THE ITALIAN LONGITUDE CONTEXT

### 3.1 Decline of the Venetian Republic

The possible existence of a Venetian longitude prize alongside prizes and monetary rewards offered by rival maritime powers—Spain, the Dutch Republic and later Britain and France—is rather curious, for a number of reasons.

First, the northern Italian city states, including Venice and Florence, were early centres of world trade. Marco Polo (1254–1324) was a thirteenth-century merchant from Venice who travelled the fabled overland Silk Road to China, while the Florentine astronomer Paolo dal Pozzo Toscanelli (1397–1482) is said to have received important insights, in 1433, from a visiting Chinese ambassador (to Florence) about the possibility of reaching China via a Western route (Wang, 2006).

At the height of its influence, in the thirteenth century, the Venetian maritime empire had established bases along the Greek coast, on Aegean Islands, in Constantinople (present-day Istanbul, Turkey) and the Levant, and along the Black Sea. Given its location on the Adriatic Coast, deep inside the northernmost arm of the Mediterranean Sea, an inland sea, Venetian merchant and navy ships did not require urgent access to accurate means to determine one's longitude at sea, out of sight of coastal landmarks. In fact, contemporary Venetian cartographers stood at the basis of the development of 'portolan' charts (e.g., Nicolai, 2016), highly accurate maps for crossing inland seas like the Mediterranean, where the effects of large-scale currents or major storms were minimal or avoidable.

Second, by the middle of the seventeenth century, Venetian industrial priorities had shifted from manufacturing goods for export to providing local services, with a focus on regional rather than long-distance markets. A range of political, economic and social changes led to a decline in the international standing of the Venetian Republic (Davids, 2013). Technical productivity started lagging and international demand for high-quality Venetian products stalled, while the Venetian government was slow in its response to the changing international order.

Rival nations, particularly those that embraced transoceanic voyages, started to thrive at the expense of once-prosperous Venice. The Venetian Republic had lost its dominant position in the increasingly globalised trade network. In this context, it is indeed curious as to why a nation in decline may have offered a significant monetary reward for an endeavour that would not immediately return dividends—and in particular an award that was clearly not widely advertised, even at the time.

**3.2 Italian longitude efforts**

Let us now consider Italian contributions to solving the longitude problem more broadly. In Section 2.2 we already encountered Galileo's evolving contributions, which were triggered by his observations of Jupiter's satellites. While Galileo was clearly on the right track, the practical application of his idea was anything but straightforward. His tables were insufficiently accurate to aid in predicting the satellites' ephemerides, and the optics of his telescopes were initially of relatively poor quality, particularly as regards the instruments' field of view. The latter aspect made timing measurements of the moons' occultations by the planet difficult and uncertain, because the small field of view rendered it difficult to find and retain the objects in one's field.

Perhaps more fundamentally, Galileo made assumptions as regards the moons' orbits, suggesting that they were circular rather than elliptical and ignoring their inclinations with respect to the plane of Jupiter's orbit. Moreover, he did not take into account the Earth's atmospheric refraction, nor the finite speed of light (which would affect eclipse timings over extended periods of time as the distance from the Earth to Jupiter would change significantly). Nevertheless, Galileo was convinced that accurate longitude determination based on Jupiter's satellites would eventually be helpful for maritime purposes, since careful, land-based mapping of the coastlines would also aid maritime navigation.

However, Galileo never managed to compile sufficiently accurate ephemerides. That was eventually achieved in 1668 by Giovanni Domenico Cassini (1625–1712; Valleriani, 2010) and marked the start of large-scale longitude measurements on land. Cassini's improved ephemerides, based on the meridian of Bologna, his home city, resulted in universal praise across the European continent. In turn, in 1669 it led to an attractive offer of a well-paid, permanent appointment at the French court of King Louis XIV (reigned 1643–1715). Cassini was subsequently appointed Director of the Paris Observatory, specifically to find the definitive longitude solution. He became a lifelong defender of Jupiter's moons as the ultimate natural clock for longitude purposes.

Meanwhile, others in Galileo's circle of friends had also become interested in contributing to a solution to the longitude problem. The Venetian mathematician Giovanni Francesco Sagredo (1571–1620) had become a minor celebrity as regards terrestrial magnetic field studies. Tycho Brahe referred to him in a letter to William Gilbert (ca. 1544–1603), the English natural philosopher, as a "… great Magneticall man" (Wilding, 2014: 33). Gilbert's most important scientific manuscript, *De Magnete, Magneticisque Corporibus, et de Magno Magnete Tellure* (*On the Magnet and Magnetic Bodies, and on the Great Magnet the Earth*; 1600), led him to correctly suggest that the Earth had an iron core and that this property determined the behaviour of the compass needle. It was hoped that these studies would eventually lead to a workable means of longitude determination using compass declinations. However, as in Spain around the same time (de Grijs, 2020a), ultimately magnetic declinations did not meet the brief of the prevailing longitude prizes and rewards.

Finally, allow me to provide a second data point on the other extremity of our timeline, simultaneously highlighting additional Italian involvement in the quest for longitude at the time. Following the announcement of the British Longitude Prize in 1714, one British projector managed to even involve Pope Benedict XIV (Prospero Lambertini; 1740–1758) in her attempts at leaving a mark. Benedict XIV was a long-standing patron of the Bologna Academy of Sciences, who was keen to facilitate reinvigoration of the sciences in Italy. He was particularly enlightened in the sense that he strongly supported talented women (Findlen, 2011). Encouraged by the pope's modern attitude, the English mathematician and astronomer Jane Squire (1671–1743) sent him copies of her treatise, *A Proposal for Discovering Our Longitude* (1742; see Figure 6), in English and French, hoping to gain his support.

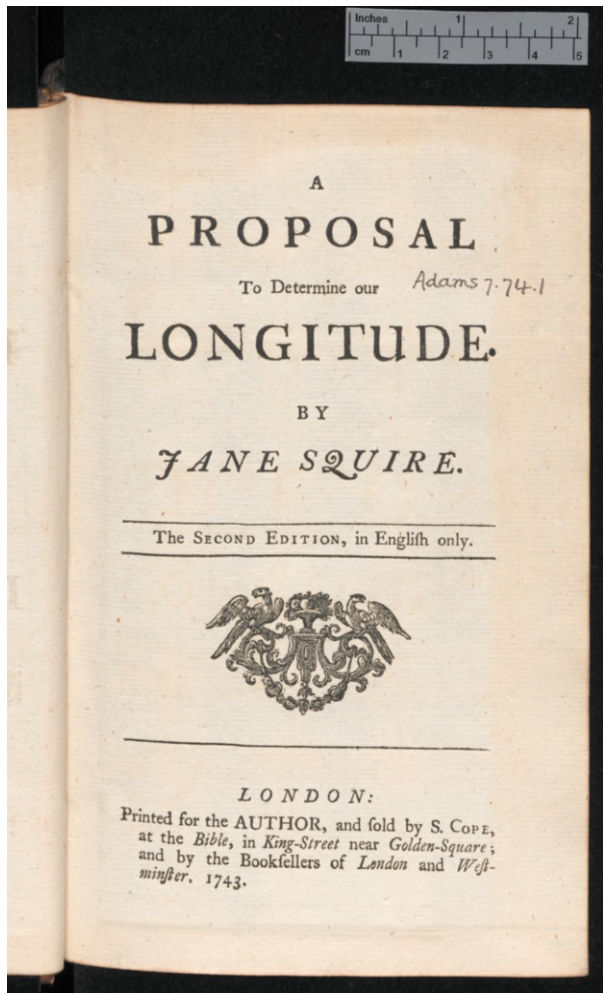

**Figure 6.** Front page of Jane Squire's treatise, *A Proposal for Discovering Our Longitude* (London, second edition, 1743). (© *Cambridge University Library; Creative Commons Attribution-NonCommercial 3.0 Unported License, CC BY-NC 3.0*).

The British scientific establishment, embodied by the Commissioners of the Longitude, had consistently ignored her advances since she first appeared on the scene in 1731 (e.g., Gingerich, 1996), which she suspected was because of her gender (Findlen, 2011; Baker, 2014, 2015). This was also acknowledged by Sir Thomas Hanmer (1677–1746), one of the Commissioners of the Longitude, in 1741: "You are to expect to lye under some Prejudice upon account of your Sex" (Baker, 2014). In 1733, she had already complained,

> I do not remember any Play-thing, that does not appear to me a mathematical Instrument; nor any mathematical Instrument, that does not appear to me a Play-thing: I see not, therefore, why I should confine myself to Needles, Cards, and Dice. (Baker, 2014)

In 1743, Pope Benedict XIV tasked the Bologna Academy of Sciences with evaluating her scheme (Benedict XIV, 1743), which required division of the sky into more than a million segments that could be recognised visually. This would allow young sailors to forego advanced mathematics in their quest to determine their longitude. Although the Academy's assessment was not encouraging (Bazzani, 1743) but rather deemed impractical, it certainly generated interest among her contemporaries:

> You must certainly have seen Mrs. Squire's scheme of the longitude, and I make no doubt understood it; but for my own part I never beheld so incomprehensible a thing in my whole life. (Elizabeth Carter, 1973; in Baker, 2014)

In fact, Squire appears to have been driven by a measure of religious fervour and advocated the use of 'absolute', Christian time. She proposed use of an 'astral' clock, which commenced longitude measurements with the zenithal star of Bethlehem at Christ's birth (Kuhn, 1984). The scheme also suggested means to correct for apparent and mean solar time using a newly invented clock (which would enable one to make celestial and terrestrial time parallel), announcing the time from church steeples using speaking trumpets and deploying artificial sea creatures as marine buoys to aid in mapping (Baker, 2014).

Moreover, and in line with Squire's public appeal to Catholicism (a questionable proposition in her Anglican environment), she promoted the development of a new universal language so as to facilitate a return to the state that existed prior to the fall of the Tower of Babel:

> May not an easy method to determine our longitude, and as easy language, whereby all mankind may understand each other, recounting the mercies of God, and the Glory of His Kingdom, be sought in a view of hastening the coming of that Kingdom? (Squire, 1742: 110)

## 4 CURRENT STATUS

In the previous sections, I have outlined the evidence for and against an alleged longitude prize offered by the Venetian Republic some time in the seventeenth century, as well as the boundary conditions and constraints available to us from historical records. On balance, I conclude that this oft-cited award was likely invented by a less-than-reliable character exhibiting signs of an inferiority complex hidden by his blustery persona.

My arguments in support of this conclusion are the following:

1. Despite frequent references to a Venetian longitude prize in modern reviews, we only have a single reference to the alleged prize ('reward') from contemporary sources (Harrison, 1696), a reference that is admittedly nothing more than hearsay.
2. Harrison's (1696) scathing assessment of the conditions governing the Venetian longitude prize simultaneously disparages the longitude reward offered by the Dutch States General, which however had long run its course by the end of the seventeenth century. This suggests that Harrison's commentary may not be an up-to-date reflection of the contemporary state of play.
3. Whereas other longitude rewards offered by the leading European maritime nations, including Castile (Spain), the Dutch Republic and Great Britain, attracted applicants and hopefuls from far and wide, often accompanied by extensive self-published pamphlets, the Venetian longitude prize does not seem to have attracted a similarly wide audience. In fact, although the archives of the Venetian senate are still being digitised, at the present time there are no digital records from the relevant period that allude to a longitude award.
4. It is curious that a city state like the Venetian Republic may have offered a longitude prize given that (i) its mariners had ready access to excellent portolan charts to travel across the Mediterranean and its neighbouring seas; (ii) historically, much of Venice's prominence in the world's trade networks came from the city's access to the Silk Road, either over land or across the Indian Ocean, but in either case this connected to the eastern Mediterranean where the longitude problem was not a major concern; and (iii) at the time of the reported Venetian award, the city's fortune was clearly in decline, and longitude determination on the open seas does not appear to have been a priority.

Instead, Harrison's 1696 reference to a Venetian longitude reward resembles the 'privileges' granted by the Senate of the Venetian Republic to 'particular persons' (S. Minuzzi, personal communication; April 2021). Henceforth, I therefore recommend that, until such a time when new and relevant information becomes available, authors refrain from referring to a Venetian longitude prize in the same context as the major sixteenth- to eighteenth-century European rewards offered by the dominant sea-faring nations.

> *I suspect that Harrison's 1696 reference to a Venetian longitude reward became part of popular history of science narratives following Turner's (1996) contribution to the 1993 longitude conference at Harvard University. That conference was also attended by Dava Sobel, whose best-selling 1995 novel,* Longitude: The True Story of a Lone Genius Who Solved the Greatest Scientific Problem of His Time*, propelled the story into the limelight. Given the wide reach of Sobel's 1995 book, I believe it likely that inclusion of a Venetian prize in her story prompted others—particularly authors outside the scholarly ecosystem—to do likewise, despite the existence of earlier references to a Venetian prize. In the context of my research for the present paper, I asked Dava Sobel if she could provide me with any additional details about a Venetian award she might have come across during her own research, but she was unable to.*

## 5 NOTES

[1] I will explain my choice of the word 'curious' in the context of the alleged Venetian longitude prize in Section 4.

[2] The anonymous author of an article published on behalf of the Horological Institute of America (1960) cites an unverified reward of 100,000 *livres* apparently offered by the Dutch government in 1716. However, he also comments that inquiries made for him by Captain (Johan Lambertus Hendrikus) Luymes (1869–1943), Netherlands Hydrographer (Head of Hydrography, 1920–1934), did not yield any information on that alleged reward. In the detailed research for my recent article on the Dutch longitude rewards (de Grijs, 2021a), I have not come across such a late Dutch reward either.

[3] In this context, 'local' applicants would be expected to come from the wider region in northern Italy, including from Padua, Florence and Bologna.

[4] The four largest moons of Jupiter were known internationally as the "Medician stars".

[5] Although little is known today about Giuliano de' Medici's life, he is included in the de' Medici's root family tree, as brother to Caterina and Cosimo de' Medici:
https://en.wikipedia.org/wiki/Medici_family_tree#Root_Medici_Tree

## 6 ACKNOWLEDGEMENTS

I gratefully acknowledge helpful discussions and email exchanges with Alexi Baker (Yale University), Paola Benussi (State Archives of Venice), Sabrina Minuzzi (Università Ca' Foscari Venezia), Simon J. Schaffer (University of Cambridge), Dava Sobel and Anthony J. Turner.

*Scientific Problem of His Time.* London, Walker Books.

Stephens, C., 2014. *Invention, Time, and Navigation: How the sea clock changed navigation and timekeeping*. Lemelson Center for the Study of Invention and Innovation, Smithsonian Institution. https://invention.si.edu/invention-time-and-navigation-0 [accessed 2 April 2021].

Squire, J., 1742. *A Proposal for Discovering Our Longitude*. London.

Turner, A.J., 1996. *In the wake of the Act, but mainly before*. In Andrewes, W.J.H. (ed.), 1996. *The Quest for Longitude*. Cambridge MA, Collection of Historical Scientific Instruments, Harvard University. Pp. 115–132.

Usher, P., 1929. *A History of Mechanical Inventions*. Ch. X. New York, McGraw-Hill.

Valleriani, M., 2010. Galileo Engineer. *Boston Studies in the Philosophy of Science*, 269. P. 59.

Wang, T.P., 2006. *The papacy and ancient China*; http://www.gavinmenzies.net/wp-content/uploads/2011/08/thepapacyandancientchina.pdf [accessed 31 March 2021].

Wilding, N., 2014. *Galileo's Idol: Gianfrancesco Sagredo and the Politics of Knowledge*. Chicago, University of Chicago Press.
13